\documentclass[manuscript]{acmart}

\usepackage{booktabs}
\usepackage{tikz}

\acmConference[IUI Companion '27]{Companion Proceedings of the 32nd International Conference on Intelligent User Interfaces}{February 8--11, 2027}{Helsinki, Finland}
\acmBooktitle{Companion Proceedings of the 32nd International Conference on Intelligent User Interfaces (IUI Companion '27)}
\acmYear{2027}

\tikzset{callout/.style={circle,fill=black,text=white,inner sep=1.2pt,font=\bfseries\scriptsize,minimum size=9pt}}

\begin{document}

\title{Glassbox: Argument Maps for Interpretable, Inspectable, and Interactive LLM Reasoning}

\author{Bart{\l}omiej Nowak}
\authornote{All authors contributed equally to this work.}
\email{nowak.bartlomiej.06@gmail.com}
\affiliation{\institution{Delft University of Technology}\city{Delft}\country{The Netherlands}}

\author{Vladimir Kirill Bickov}
\authornotemark[1]
\email{me@nevolodia.com}
\affiliation{\institution{Delft University of Technology}\city{Delft}\country{The Netherlands}}

\author{Egor Georgievskii}
\authornotemark[1]
\email{egeorgievskii@gmail.com}
\affiliation{\institution{Delft University of Technology}\city{Delft}\country{The Netherlands}}

\author{Ioannis Pantidis}
\authornotemark[1]
\email{pantidisi@gmail.com}
\affiliation{\institution{Delft University of Technology}\city{Delft}\country{The Netherlands}}

\author{Fiodar Prystauka}
\authornotemark[1]
\email{fiodar.prystauka@gmail.com}
\affiliation{\institution{Delft University of Technology}\city{Delft}\country{The Netherlands}}

\renewcommand{\shortauthors}{Nowak et al.}

\begin{abstract}
  Large language models (LLMs) often give confident answers backed by reasoning that is flawed or unfaithful to their decision process. Chain-of-thought (CoT) prompting exposes this reasoning, but only as unstructured free text that is hard to analyze systematically and offers no guarantee of faithfulness. We present Glassbox, a chat system that pairs each LLM answer with an argument map, making the model's stated reasoning interpretable, inspectable, and interactive. Each map is a typed graph of claims and their relations, built in one of four argumentation frameworks. While the answer streams, an ensemble of LLM judges assesses each map and the generating model repairs any that fail, with every verdict and its feedback visible to users. For quantitative (QBAF) and Bayesian maps, users can also contest the reasoning by adjusting, adding, or removing nodes. Across three benchmarks and two models, reasoning through argument maps incurs no systematic loss of accuracy relative to CoT. Glassbox thus makes LLM reasoning structured, auditable, and contestable without sacrificing performance.
\end{abstract}



\maketitle

\section{Introduction and Background}

Large language models (LLMs) are increasingly deployed in high-stakes domains such as finance, healthcare, and law~\cite{chen2024a}, where explanations must faithfully reflect the reasoning behind an answer~\cite{explainability}. Chain-of-thought (CoT) prompting~\cite{wei2022chain} exposes an LLM's reasoning by having it write out intermediate steps, yet such explanations can systematically misrepresent the true reason for a model's prediction~\cite{turpin2023language}. Moreover, because CoT is unstructured free text, readers must first identify the claims and reconstruct how they relate before they can verify them. Argument maps~\cite{argmapsGeneral} remove this burden by making claims and their relations explicit. They have also been shown to improve comprehension~\cite{dwyer2013examination} and critical thinking~\cite{criticalthinking}, making them a natural format for reasoning that people must understand and check. This has motivated work on argumentative LLMs. Freedman et al.~\cite{Freedman_2025qbaf} build on quantitative bipolar argumentation frameworks (QBAFs)~\cite{qbaf}: an LLM generates and scores arguments for and against a claim, and DF-QuAD~\cite{dfquad} computes the claim's strength. They formalize contestability and evaluate it on binary claim verification. ArgLLM-App~\cite{dejl2026argllmapp} makes this approach interactive: users pose a binary claim and contest the outcome by changing confidences or adding arguments.

This demo presents Glassbox, a chat interface that makes an LLM's stated reasoning \emph{interpretable}, by expressing it as a typed argument map rather than free text; \emph{inspectable}, by having an LLM-as-a-judge~\cite{zheng2023judging} ensemble assess every map and showing users its verdicts and feedback; and \emph{interactive}, by letting users contest the reasoning through map edits to which the model must respond. Unlike ArgLLM-App, Glassbox handles open-ended chat rather than binary claims, chooses among four argumentation frameworks per question instead of QBAFs alone, and judges every argument map it produces.

\section{System Description}

\begin{figure}[tbp]
\centering
\begin{tikzpicture}
\node[anchor=south west,inner sep=0] (img) at (0,0) {\includegraphics[width=0.8\linewidth]{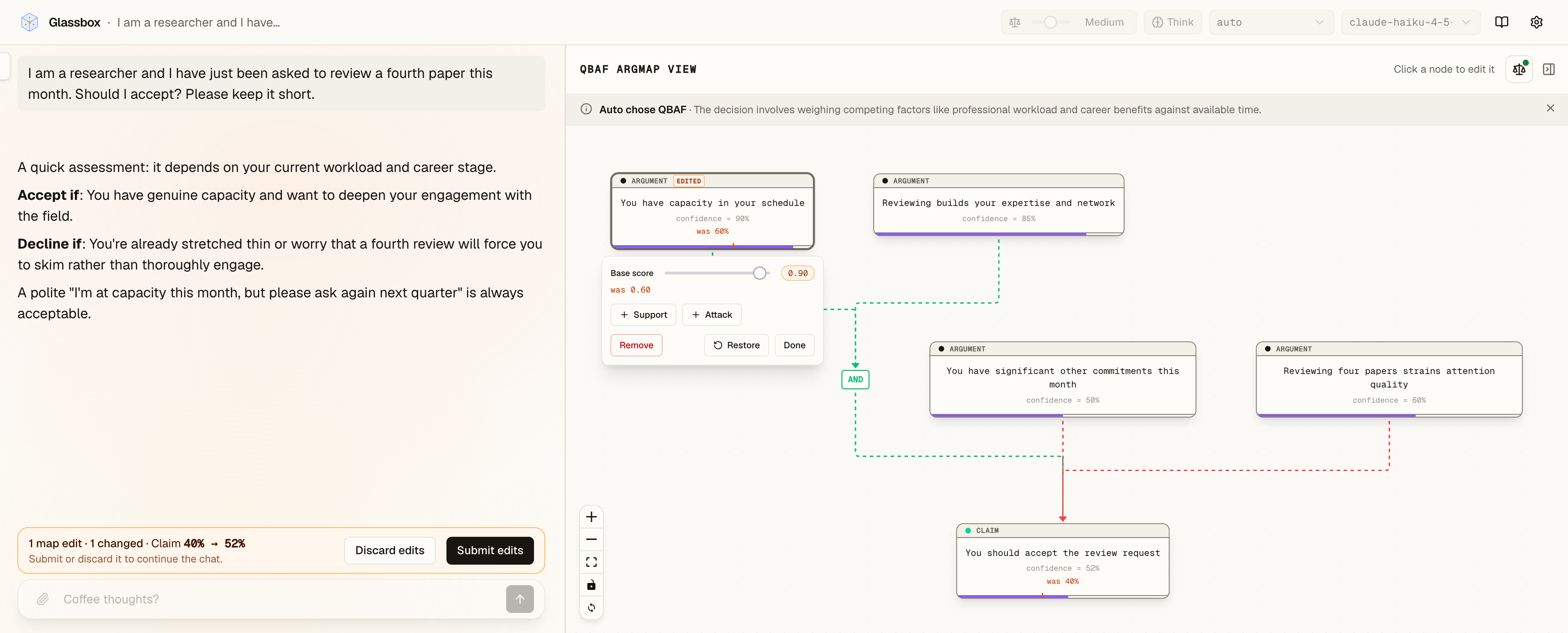}};
\begin{scope}[x={(img.south east)},y={(img.north west)}]
  \node[callout] at (0.353,0.828) {A};
  \node[callout] at (0.372,0.750) {B};
  \node[callout] at (0.969,0.830) {C};
  \node[callout] at (0.378,0.666) {D};
  \node[callout] at (0.376,0.505) {E};
  \node[callout] at (0.6025,0.114) {F};
  \node[callout] at (0.011,0.190) {G};
  \node[callout] at (0.623,0.965) {H};
\end{scope}
\end{tikzpicture}
\Description{Screenshot of the Glassbox interface. The left pane shows a chat with a question about whether to accept a fourth review request this month, the model's short reply, and a banner above the message box that reads one map edit, claim 40 percent to 52 percent, with Discard edits and Submit edits buttons. The right pane shows a QBAF argument map with five nodes. The evidence node You have capacity in your schedule is marked edited and has a small editor open below it with a confidence slider and Support, Attack, Remove, Restore, and Done buttons. A line above the map reads Auto chose QBAF followed by a one-sentence reason. Eight letters A to H mark the elements described in the caption.}
\caption{The Glassbox interface while a user contests an argument map. (A)~Auto mode's chosen framework and rationale. (B)~The argument map (here a QBAF), with strengths computed by CLU. (C)~The judging-view button (Figure~\ref{fig:views}a). (D)~An edited node with its previous value. (E)~The node editor: confidence slider, add-argument buttons, and Remove; derived steps can only be removed, and the claim is locked. (F)~The claim's new and previous values. (G)~The message box, locked while edits are pending. (H)~The judge-level slider.}
\label{fig:interface}
\end{figure}

Glassbox is a web chat application\footnote{Available at \url{https://project-glassbox.com}.} that displays the argument map of the selected answer alongside the conversation (Figure~\ref{fig:interface}). A React frontend communicates with a Go backend, which streams replies from the model runtime and uses a core logic unit (CLU) to parse each argument map into typed JSON with a grammar.

\subsection{Interpretable: Reasoning as a Typed Argument Map}
Whereas CoT expresses reasoning as free text, Glassbox represents it as a typed graph (Figure~\ref{fig:interface}, B) of claims (nodes) and their relations (edges). Four frameworks are supported: propositional maps, whose formulas are linked by named inference rules~\cite{whately1854elements}; Bayesian maps, with priors and derivations~\cite{jensen2001bayesian}; QBAFs, with confidences and support or attack edges~\cite{qbaf}; and Toulmin maps, with claim, grounds, warrant, backing, and rebuttal~\cite{toulmin2003uses}. Propositional and Toulmin maps carry no numeric strengths and are read from their structure alone, whereas CLU computes strengths for QBAF and Bayesian maps. In auto mode, a selector picks a framework, or none, and states its rationale (Figure~\ref{fig:interface}, A).

\subsection{Inspectable: LLM-as-a-Judge and the Judging View}
\begin{figure}[tbp]
\centering
\includegraphics[width=0.9\linewidth]{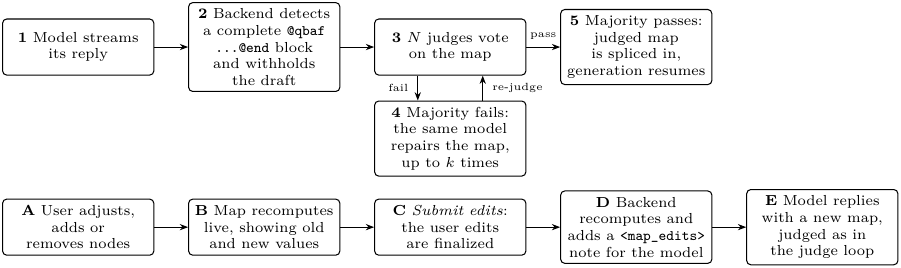}
\Description{Diagram of two loops. The judge loop: the model streams a reply, the backend detects a complete map block and withholds the draft, a set of judges votes on the map; on a pass the judged map is spliced in and generation resumes, on a fail the same model repairs the map and the judges vote again. Below it, the contest loop: the user adjusts, adds, or removes nodes, the map recomputes, the backend recomputes and adds a note for the model, and the model replies with a new map that is judged as above.}
\caption{The two loops. The judge loop (top) checks and repairs each argument map during streaming (shown here for a QBAF block); the contest loop (bottom) feeds user edits back to the model, whose new map is judged again.}
\label{fig:architecture}
\end{figure}

A slider (Figure~\ref{fig:interface}, H) selects the judge level (none, low, medium, high, max), which sets the number of judges and the maximum number of repairs to \mbox{$(N,k)=(0,0)$}, $(1,1)$, $(3,2)$, $(5,2)$, and $(7,3)$, respectively; the demonstration uses medium. Judging runs while the answer streams (Figure~\ref{fig:architecture}): once a map block is complete, the backend pauses generation, withholds the draft, and sends the argument map to the judges. Each judge votes pass or fail on a rubric of syntax, validity, readability, contrastivity, completeness, and calibration, and a map passes on a strict majority. Otherwise, the generating model repairs the map using the judges' feedback, and the repair is judged again; if all $k$ repairs fail, the final map is still shown. The judging view (Figure~\ref{fig:interface}, C; Figure~\ref{fig:views}a) lists each run's votes, map, and judge assessments.

\subsection{Interactive: Contesting the Reasoning}

\begin{figure}[tbp]
\centering
\begin{minipage}[t]{0.48\linewidth}
  \centering
  \includegraphics[width=0.9\linewidth]{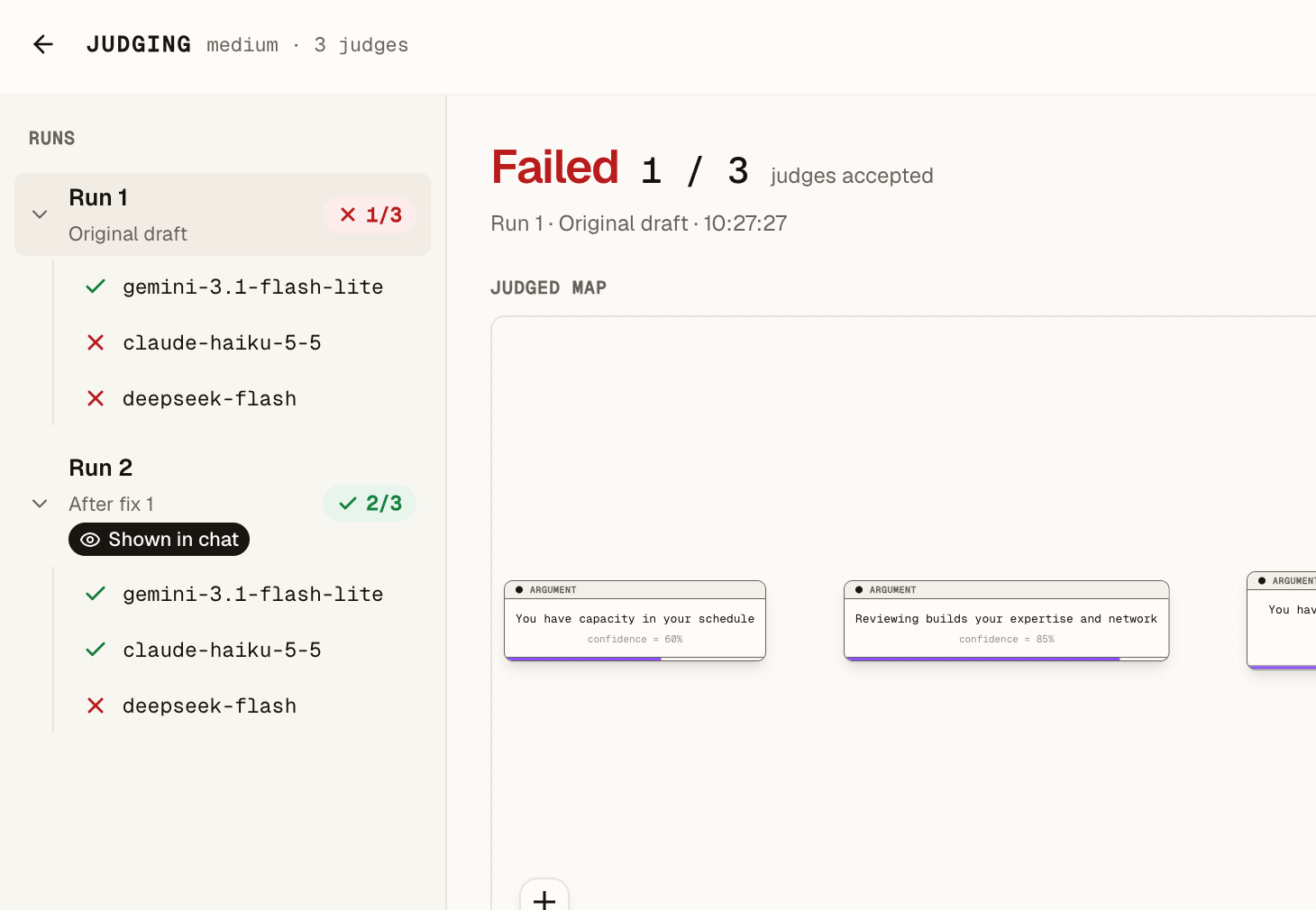}\\[2pt]
  \small\textbf{(a)}
\end{minipage}\hfill
\begin{minipage}[t]{0.48\linewidth}
  \centering
  \includegraphics[width=0.9\linewidth]{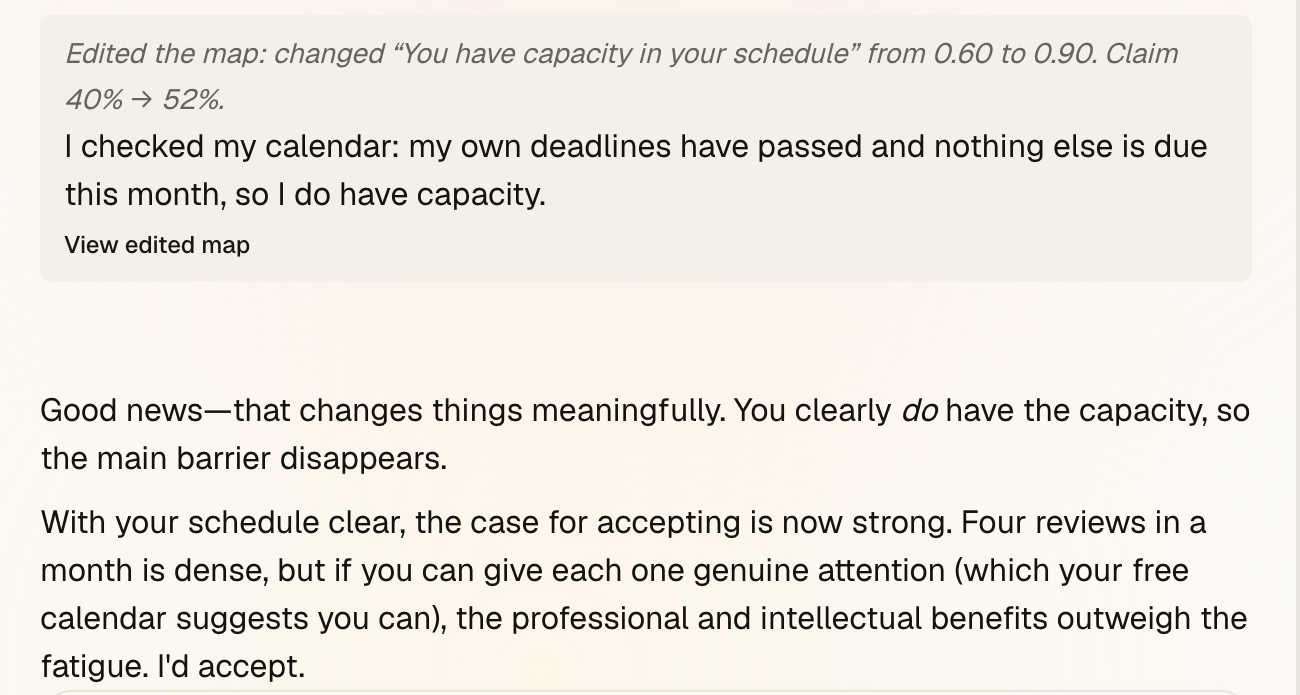}\\[2pt]
  \small\textbf{(b)}
\end{minipage}
\Description{Two screenshots. (a) The judging view: two judging runs listed on the left, the original draft of the map rejected by two of three judges and the repaired map accepted by two of three and marked as shown in chat, and on the right the first judged map, in which two supporting nodes are not connected to the claim. (b) The chat after contesting: an italic line reading Edited the map, changed You have capacity in your schedule from 0.60 to 0.90, claim 40 percent to 52 percent, followed by the user's note and a View edited map link, then the model's reply that adopts the edit.}
\caption{(a)~The judging view for the scenario in Section~\ref{sec:scenario}: the rejected first draft and the repaired run (generator Claude Haiku 4.5; judges Gemini 3.1 Flash-Lite, Claude Haiku 5.5, DeepSeek V4.1 Flash). (b)~The chat after contesting: edit summary, note, and reply.}
\label{fig:views}
\end{figure}

Users can contest the latest QBAF or Bayesian map (Figure~\ref{fig:interface}, D--G; Figure~\ref{fig:views}b) in three ways: by adjusting the evidence (the prior in Bayesian maps, the confidence in QBAFs), by adding their own argument node, or by removing an existing one. The argument map recomputes immediately, showing each edited node's previous value and the resulting shift in the claim. Changes remain reversible until the user presses \emph{Submit edits}; the model then receives a note listing them alongside all recomputed values, and is instructed either to adopt the edits or to justify retaining its original map.

\section{Demonstration Scenario}
\label{sec:scenario}

\textbf{Step 1: Ask.} With auto mode and judge level medium, a researcher asks whether to accept a fourth review request this month. Auto mode selects QBAF because the decision weighs workload and career benefits against available time (Figure~\ref{fig:interface}, A).

\noindent\textbf{Step 2: Inspect.} Two of three judges reject the first draft (Figure~\ref{fig:views}a) because its two supporting nodes are disconnected from the claim, leaving only attacks. The repaired map, accepted by two of three judges, rates the claim at 40\%.

\noindent\textbf{Step 3: Contest.} Without rephrasing the question, the researcher raises the confidence of the capacity node from 0.60 to 0.90, watches the claim rise to 52\% (Figure~\ref{fig:interface}, D--F), and submits the edit with a note.

\noindent\textbf{Step 4: Compare.} The model adopts the edit and drops the ``other commitments'' attack the note resolves (Figure~\ref{fig:views}b); two of three judges accept the new argument map. Compared with plain chat, the researcher has seen why the model leaned toward declining, seen a flawed map repaired, and raised the claim from 40\% to 63\% by correcting one assumption.

\section{Evaluation and Future Work}

\textbf{Accuracy.} Argument maps are valuable only if they do not degrade the answers they accompany, so we tested whether they reduce accuracy. With judging disabled to isolate the effect of the maps, we compared each framework and auto mode against CoT for DeepSeek V4.1 Flash and Claude Haiku 5.5 on HellaSwag~\cite{zellers-etal-2019-hellaswag}, the multiple-choice subset of HLE~\cite{hle}, and StrategyQA~\cite{strategy_qa} (Table~\ref{tab:accuracy}). Maps stay close to CoT accuracy without systematic degradation, and some frameworks score slightly higher, e.g., QBAF by 1.4 points on HellaSwag (DeepSeek) and Bayesian by 3.0 points on HLE (Claude). The largest drops are on StrategyQA, where Toulmin and QBAF trail CoT by about 5 points for DeepSeek.

\begin{table}[htbp]
\footnotesize
\caption{Accuracy (\%) with no map (CoT), each of the four frameworks, and auto mode; question counts in parentheses. HLE-MC is the multiple-choice subset of HLE. Best per row in bold; less than 1 point below the best underlined.}
\label{tab:accuracy}
\begin{tabular}{llcccccc}
\toprule
Model & Benchmark & CoT & Prop. & Toulmin & QBAF & Bayes. & Auto \\
\midrule
DeepSeek V4.1 Flash & HellaSwag (1000) & 90.0 & 88.0 & \underline{90.9} & \textbf{91.4} & 86.5 & 89.8 \\
 & HLE-MC (563) & \underline{20.2} & 18.3 & \textbf{20.6} & \underline{19.7} & 16.2 & 18.3 \\
 & StrategyQA (682) & \underline{84.6} & 84.0 & 79.9 & 79.5 & \textbf{85.5} & 83.6 \\
\midrule
Claude Haiku 5.5 & HellaSwag (1000) & \underline{85.8} & \underline{86.2} & \underline{86.0} & \underline{86.2} & \underline{85.8} & \textbf{86.5} \\
 & HLE-MC (563) & 25.6 & 26.5 & 25.6 & 26.6 & \textbf{28.6} & 26.3 \\
 & StrategyQA (682) & \textbf{84.6} & \underline{84.3} & \underline{83.7} & 82.8 & \underline{84.3} & 83.6 \\
\bottomrule
\end{tabular}
\end{table}

\textbf{Limitations and next steps.} An argument map renders the model's stated reasoning, not its internal computation, and the judges have not yet been validated against human raters. We therefore plan to measure judge--human agreement, test map faithfulness under perturbation, and study whether maps help people detect injected errors.


\section*{GenAI Usage Disclosure}
Claude (Anthropic) assisted with the system code and this paper's phrasing; the authors take responsibility for both.

\bibliographystyle{ACM-Reference-Format}
\bibliography{references}

\end{document}